# The effect of constraints on information loss and risk for clustering and modification based graph anonymization methods

David F. Nettleton[1,2], Vicenc Torra[2], and Anton Dries[3]

[1]Universitat Pompeu Fabra, Barcelona, Spain; [2]IIIA-CSIC, Bellaterra, Spain; [3]KU Leuven, Leuven, Belgium.

**Abstract**—In this paper we present a novel approach for anonymizing Online Social Network graphs which can be used in conjunction with existing perturbation approaches such as clustering and modification. The main insight of this paper is that by imposing additional constraints on which nodes can be selected we can reduce the information loss with respect to key structural metrics, while maintaining an acceptable risk. We present and evaluate two constraints, 'local1' and 'local2' which select the most similar subgraphs within the same community while excluding some key structural nodes. To this end, we introduce a novel distance metric based on local subgraph characteristics and which is calibrated using an isomorphism matcher. Empirical testing is conducted with three real OSN datasets, six information loss measures, five adversary queries as risk measures, and different levels of $k$-anonymity. The result show that overall, the methods with constraints give the best results for information loss and risk of disclosure.

**Index Terms**— privacy, information hiding, graphs and networks

## 1 INTRODUCTION

Data Privacy in graphs has recently become a topic of renewed interest by researchers, partially due to the emergence of online social networks (OSN), which can be represented and analyzed as graphs. OSN data is of great potential for data analysts from different disciplines, but also represents a threat to data privacy if it is used for the wrong motives. However, anonymization of this type of data represents a challenge, given that anonymization techniques may destroy or impair essential structural information in the graph.

The objective of the current work is to test different perturbation and selection methods under different conditions to evaluate their relative performance.

In the literature, some authors have considered anonymization as a graph partitioning/clustering task based on an overall utility measure[1] or by modifying nodes using a cost function[2]. However, there is less work done to benchmark these distinct methods together and under restricted conditions. Also, the high cost and/or complexity of graph search based on isomorphic properties, has led to heuristic distance based approximate matching methods. The authors propose that the latter methods will be appropriate when the adversary knowledge is based on graph topological queries, such as in the present paper, and/or when the perturbation method is based on clustering in which case an adversary isomorphic query on the local subgraph is ineffective.

**The structure of the paper is as follows:** in Section 2 we discuss the state of the art for the issues we are considering in this paper; in Section 3 we present preliminary concepts and discuss some aspects such as risk due to information leak, the privacy model, the merging process and information loss; in Section 4 we describe what is understood as a local neighborhood subgraph, define the distance metric, graph alteration operators, restrictions, and finally give pseudo-code for the method; in Section 5 we define the metrics used for information loss and adversary knowledge, and define the privacy model; in Section 6 we describe the test datasets and experimental procedure; in Section 7 we present the empirical results for infor-mation loss and risk of disclosure (adversary queries) for three distinct graph datasets, five different perturbation methods and variants, and different values of the privacy level *k*; finally, in Section 8 we present the summary and some conclusions.

## 2 RELATED WORK

The theme of privacy preserving social network publishing is considered from two general perspectives: (i) adversary information and (ii) anonymization methods.

### 2.1 Adversary Information

Adversary information is a way of evaluating the risk of reidentification and normally involves 'attacking' the data with informational queries which take into account the type and amount of knowledge available to the adversary. In [1], Hay et al. consider what an adversary may know or deduce from a graph in terms of three different families of topological queries (as opposed to isomorphic properties). In general, the queries focus on eliciting information about the immediate or close neighborhood of a target node. The first query is called *'vertex refinement'*, which returns the degree of a target node, of its immediate neighbors, and so on. The second query is called a *'subgraph query'* and returns the number of edges of a given neighborhood subgraph. The third query is called a *'hub fingerprint'*, which returns information about the proximity of a target node to one or more hub nodes in the graph. We enter into more detail about these three types of information in Section 5.3, given that we use them to evaluate risk of disclosure in the present work. Wondracek [3] presented a different approach, in that the attacker uses a malicious website to obtain information about users of an on-line social network. Backstrom et al. [4] is a key reference for adversary strategies, which are divided into active attacks, in which the adversary actively tries to affect the data to make it easier to decipher, and passive attacks, in which the adversary simply observes data as it is presented. The graph attacked is considered to be a single anonymized copy of a social network, and the adversary queries consider isomorphic properties. In [5], Cheng et al. consider a K-Isomorphism approach to privacy preserving network publication which protects against structural attacks. The authors refer to a popular type of attack described by Backstrom et al. in [4], which involves the use of embedded subgraphs. They extend this idea by defining two realistic attack targets which they call 'NodeInfo' and 'LinkInfo'. They show that k-isomorphism, or anonymization by forming k pair wise isomorphic subgraphs, is both sufficient and necessary for protection. However, the process is shown to be NP-hard. They present some techniques which enhance the anonymization efficiency while retaining the data utility.

### 2.2 Anonymization Methods

In the literature, different methods have been used for graph anonymization and in particular, obtaining k-anonymity of the vertices V in a graph G, while minimizing information loss. For the purposes of the current work we will divide the methods into two groups (as has been done by several authors in the literature): (a) node modification approaches and (b) node clustering approaches. In the context of data privacy in general, Sweeney's paper [6] was the first to define *k*-anonymity, and more recently the paper by De Capitani et al. [7], gave key definitions for privacy levels, information loss and risk of disclosure. In [8] Zhou considered l-diversity together with k-anonymity to give a stronger anonymity guarantee.

#### 2.2.1 Node modification approaches

Node modification approaches act by choosing similar nodes and making them identical. This can be done by adding nodes to make their degrees the same and by adding edges to make their immediate neighborhood connectivity the same. Using this method, k-anonymity is achieved by obtaining that every node in the graph has at least k-1 other nodes which are indistinguishable from it. Zhou[2] presents a method which selects nodes based on a cost function and then anonymizes them by adding nodes and edges to their neighborhoods. As well as anonymizing the topology, it generalizes the vertex labels. The topological aspect re-lies on an elaborate coding scheme to speed up isomorphic comparisons between subgraphs. In [9], Nettleton et al. compare two different types of online social network from a data privacy perspective, using 'add link' as the perturbation operator. In [10], Hay et al. presented a simple graph anonymization based on random addition and deletion of edges. The attack method attempts re-identification using two types of queries, vertex refinement and subgraph knowledge. The risk measure is considered as the percentage of nodes whose equivalent candidate set falls into one of a given set of buckets (1 node, 2-4 nodes, 5-10 nodes, ...). In [11], Das et al. present a linear programming-based technique for anonymization of edge weights by scrambling the relative ordering of the edges sorted by weights, which preserves the linear properties of the graph.

### 2.2.2 Node clustering approaches

Node clustering approaches act by choosing similar nodes and physically grouping them. This can be done by a k-means type algorithm or by a similarity/distance metric to choose similar nodes. Using this method, k-anonymity is achieved by obtaining that every node in the graph is incorporated into a cluster within which there are at least k-1 other nodes.

Skarkala et al. [12] and Liu and Yang [13] have recently followed similar approaches for node clustering/grouping which take into consideration the privacy protection of the edge weights. Skarkala's method [12] employs a similarity function to form clusters each containing at least k nodes. Liu's method [13], on the other hand, used a k-means type clustering by calculating cluster centers.

Nettleton in [14] applied a perturbation method based on node aggregation and a similarity metric with fixed weights for choosing node pairs. Different types of clustering, fuzzy (fuzzy c-Means) and crisp (k-Means) are applied to graph statistical data in order to evaluate the information loss due to perturbation. In [15], Nettleton and Torra define a set of six graph alteration operations for perturbation, and evaluate how each operation affects the graph. The graph alteration operations considered were 'add/delete' edge, 'add/delete' vertex and 'aggregate/disaggregate' node.

In [1], Hay presented an approach in which nodes are grouped into partitions based on a utility function incorporating a distance metric in terms of the number of edges. In order to settle the partitions, the entropy is calculated for the entire graph. However, this process may incur a high computational cost, depending on the graph size, topology and characteristics. Hay's method [1] is distinct to our approach given that Hay's partitions are guaranteed as having at least $k$ nodes but can have many more (e.g. hundreds, for $k$=16), whereas our method guarantees between $k$ and $2k-1$ nodes in each cluster.

### 2.2.3 Other approaches

In [16], Bonchi et al. offer a somewhat different vision of graph anonymization, distinguishing an entropy-based quantification of anonymity, which they consider a global method, from a local quantification based on a-posteriori belief. They also propose a controlled random removal edge (as opposed to adding edges) which they call 'random sparsification'. In [17], Ying and Wu present a spectrum preserving approach to randomizing social networks. The authors base their approach in the observation that many graph structures have a strong association with the spectrum, hence the idea to define a perturbation strategy which minimizes the change in some given eigenvalues, while maintaining privacy protection. Zou et al. [18] propose k-automorphism as an anonymity property to protect against multiple structural attacks and develop an algorithm that ensures k-automorphism.

## 3 PRELIMINARIES

A graph is defined as a set of vertices V interconnected by a set of edges, thus G = (V, E). In the current work each node has an arbitrary identifier for data processing purposes however we assume this identifier will have no meaning for the adversary and cannot be considered a label. Hence, we are dealing with an unlabeled graph. A local neighborhood subgraph $G^n$ = (V', E') is a subset of G around a given reference node $v^r$ at one hop. Hence $v^r \in V'$ and all other vertices v' ∈ V' are immediate neighbors of $v^r$.

In G we define two special sets of vertex types: hubs and bridges. A *hub* vertex $v^h$ is defined as being a node with a relatively high number of direct connections to other nodes, as quantified by Kleinberg's metric [19] which we designate as *h(v)*. We define the set of hub vertices as $V^h \subset V$, and $v^h \in V^h$ when $h(v^h)$ is in the top 12% percentile of all values for *h(v)*. A *bridge* vertex $v^b$ is defined as being a node which has a relatively high number of critical paths which go through it to/ from other nodes in the graph, as quantified by Hwang's metric [20] which we designate as *b(v)*. We define the set of bridge vertices as $V^b \subset V$, and $v^b \in V^b$ when $b(v^h)$ is in the top 10% percentile of all values for *b(v)*. The top percentile values for hubs and bridges were chosen by empirical study of the respective metric distributions.

Also we map a partitioning on G derived by the community structure identified by the Louvain Method [21]. The mapping of the vertices into the community structure can be defined as a function $G_c : v_i \to c$. Hence, a given vertex $v_i$ will belong to one and only one community *c*.

The anonymization method chooses pairs of nodes ($v_i$, $v_j$), based on a distance function $D(v_i, v_j)$ and subject to the following restrictions: $v_i \notin V^h$, $v_i \notin V^b$, $v_j \notin V^h$, $v_j \notin V^b$, $G_c(v_i) = G_c(v_j)$. These three definitions apply the hub, bridge and community restrictions, respectively.

**Possible information leaks:** the exclusion of the hub and bridge vertices and the limitation to the same community may have information leak consequences to the adversary which are compensated by the reduction in information loss. We assume that we are interested in protecting these three aspects because the user/ analyst is specifically interested in them from a utility point of view.

Consider the situation in which a given hub vertex $v^h$ with $v^e$ edges is excluded from anonymization. As we commented previously, any vertex is the reference node $v^r$ of its corresponding local neighborhood subgraph $G^n \subset G$. However, the neighbors of hub (and bridge nodes), which are not themselves hub or bridge vertices, may be anonymized. Consider the case when a neighbor of $v^h$ is also neighbor of a vertex which is neither a hub or a bridge vertex in graph G. In this case it will be a candidate for anonymization. Once the neighbor has been anonymized (that is, aggregated with another node) the resulting 'supernode' will continue having a link to the hub node (which has not been anonymized). However, when the anonymization process has been completed for k-anonymity, the hub node, which originally had N neighbors will now have N/ *k* neighbors. Also each neighbor will be a supernode containing at least *k* original nodes. Hence, although the adversary may be able to identify a given hub node, s/ he will not be able to distinguish between the *k* original nodes which comprise each supernode which is a neighbor $v^h$. Another possible risk is when a given community contains a hub node which connects to all (or a high percentage) of the nodes in that community. As mentioned previously, the neighbors of a hub node which are not themselves hubs (or bridges) can be anonymized. Hence the community could be anonymized to k-anonymity, with the exception of the hub node.

*Community size guarantee:* the Louvain method[21] partitions the complete graph into communities. However, given the restriction that nodes can only be anonymized using other nodes from the same community, we have to guarantee that there are at least *k* eligible nodes in each community, where *k* is the desired anonymity level This is achieved by using a higher value for the 'resolution' parameter of the algorithm (as implemented in Gephi version 0.8.2-beta) which controls the community size.

Hence the true potential information leak for a hub vertex $v^h$ will be $(| v^e+1| - | v^{e'}|) / N$, where $v^{e'}$ represents the vertices neighbors of $v^h$ which may be anonymized with non-hub neighbors. The same reasoning also applies for bridge nodes. A worst case scenario would be a community consisting of exactly k nodes of which a high proportion are either hubs or bridges. However, by observation, we have seen that in practice the number of hubs and bridges in a community tends to be small in relation to the total number of nodes in the community. We also note that the number of hubs and bridges excluded from anonymization is reduced by selecting only a top percentile of them. This selection process will be explained in detail in Section 4.4 of the paper.

In the case of the bridge nodes, these will tend to have a significantly lower degree than the hub nodes and therefore the number of potentially affected nodes (immediate neighbors) will be proportionately lower.

In terms of information leak probability, we consider two aspects: (i) that the number of top percentile hub and bridge vertices is small with respect to the number of vertices in the total graph, designated by $(|H| + |B|) / N$; and (ii) that in a given community, the number of neighbor nodes $v_i$ of the hub and bridge vertices ($v^h$ and $v^b$, respectively) which cannot be matched with at least one other non hub or bridge vertex in the same community is also small, designed by $|(v_i \in (V^h | V^b))| / |G^n|$, where $v_i$ is a given node in the immediate local neighborhood subgraph $G^n$ of a given hub node $v^h$ or a bridge node $v^b$. Hence the information leak probability will in practice be small and equal to: $((|H| + |B|) / N) \times (|(v_i \in (V^h | V^b))| / |G^n|)$.

The limitation that vertex pairs to be anonymized must be in the same community also has information leak considerations. The size of the community is a factor in this consideration. If we have no restriction based on community, then all vertices are candidates. If we have M communities with $M_1 N, M_2 N, \ldots M_n N$ vertices respectively, we have a corresponding reduction in candidate diversity. Also the adversary may be able to calculate the communities given that the Louvain method is a public algorithm, and reduce the search space for victim vertices. We can formalize the average reduction in candidate diversity as follows: let $N_c$ be the average number of vertices per community; then $N / N_c$ will be the reduction in diversity with respect to the complete graph, due to the partitioning into communities, which represents the potential information leak. As mentioned previously, we can guarantee that there are at least k nodes per community, using the "resolution" parameter of the Louvain algorithm.

**Privacy guarantee/model (see also Sec. 5.2):** if all N nodes are anonymized we will obtain k-anonymity. Consider a graph G which represents a social network and G' which represents an anonymized version of G. If G' is k-anonymous then an adversary who launches structural queries cannot re-identify a given vertex from G in G' with a confidence greater than 1/k. In the case of the clustering methods, each super-node in G' will contain at least k nodes from G, hence implementing k-anonymity. However, we take into consideration the aspects explained previously in this section with regard to the nodes (top percentile hubs and bridges) which are excluded from anonymization and the restriction that the node pairs to be merged are in the same community.

We highlight that in the present work we employ an anonymity model for social networks such that a graph satisfies k-candidate anonymity if for every structural query over the graph, there exist at least k nodes that match the query (the structural queries are defined in Section 5.3). This is similar to the privacy model of Hay et al. [1] and distinct from Backstrom's isomorphic anonymity [4].

**Nodes excluded from k-anonymization:** consider the extreme case, represented by a k value sufficiently large so that all nodes would be iteratively paired until no more pairings are possible in the anonymized graph G'. Given our restrictions for pairing, we would have just one supernode in each community in G', together with the hub and bridge nodes excluded from pairing. We designate the hub threshold as $\theta^h$ and the bridge threshold as $\theta^b$. That is, all hubs whose hub metric is greater or equal to $\theta^h$ will be included in $V^h$. Likewise, all bridges whose bridge metric is greater or equal to $\theta^b$ will be included in $V^b$. The thresholds are assigned empirically, as explained in Section 4.4. The number of hub nodes not anonymized would be $|\{V^h\}|$ and the number of bridge nodes not anonymized would be $|\{V^b\}|$, where $V^h$ and $V^b$ represent the sets of non-anonymized hubs and bridges in G', respectively. The neighbors of the hubs and bridges would be affected as we have commented previously in this section.

**Merging process:** the merging process is defined as follows: Two vertices $v_1$ and $v_2$ are selected for merging using a similarity function. The set of immediate neighbors of node $v_1$ is defined as $V^1 = \{v_{11}, \ldots v_{1n}\}$ with a corresponding set of links $E^1 = \{e_{11}, \ldots e_{1n}\}$. Likewise, the set of immediate neighbors of node $v_2$ is defined as $V^2 = \{v_{21}, \ldots v_{2m}\}$ with a corresponding set of links $E^2 = \{e_{21}, \ldots e_{2m}\}$. Now, consider that nodes $v_1$ and $v_2$ are merged to form a new 'super' node, $v_3$. The new set of neighbors $V^3$ and edges $E^3$ will be the union of sets $V^1$ with $V^2$ and $E^1$ with $E^2$, respectively. In the case that $v_1$ and $v_2$ have one or more common neighbors, then the resulting set will have correspondingly less edges and vertices than the sum of the original sets of edges and vertices. The merging process continues until all eligible nodes have been merged into supernodes each containing $k$ nodes.

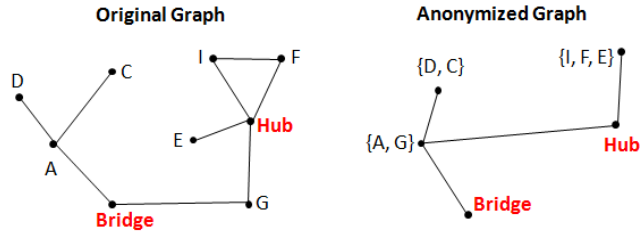

Fig. 1. Merging Process

In Fig. 1 we see a simple example of the merging process applied to an initial graph consisting of 7 simple nodes, one bridge node and one hub node. In the anonymized graph we see that the Hub and Bridge nodes have not been aggregated, whereas the remaining 7 nodes have been aggregated into 3 supernodes containing 2, 2 and 3 original nodes, respectively. Hence, this graph has a minimum anonymity of k=2 for the basic nodes. The basic nodes have been paired successively based on a similarity measure in order to minimize the overall perturbation.

With respect to the average degree, in general equilibrium is maintained, given that, on the one hand the supernodes have more links than the basic nodes, but on the other hand, there are progressively less nodes to link to (determined by the privacy value 'k'). For example, with reference to Fig. 1, in the original graph (on the left) the degree has an average of 2.0 and a standard deviation of 0.89, whereas in the k=2 anonymized graph on the right the corresponding values are 1.6 and 1.0, respectively.

**Information loss:** information loss is mitigated on two levels: **(i)** we consider a higher level in which we protect the community structure of the graph, and the top percentile hubs and bridges as key elements of the graph structure; **(ii)** at a lower level we consider a cost function which chooses nodes for clustering/ modification which minimize the perturbation to the corresponding local sub-graphs. The distance/ cost function and the perturbation methods are described in Section 4. In order to quantify the information loss at both levels we use six metrics, as described in Section 5.1: degree, clustering coefficient, average path length, hub value, bridging centrality value and number of communities, designated as metrics $m_1$ to $m_6$, respectively. In general, if $G$ is the original graph, $G'$ the perturbed graph, $m_1$ represents the degree values for the original graph, $m_1'$ represents the degree values for the perturbed graph, then the information loss will be:

**IL**$(G, G', m_1) = 1 - $ **corr**$(m_1, m_1')$     (1)

The information loss for metrics $m_2$ to $m_5$ would follow in a similar manner. In the case of $m_6$, we take the absolute difference between the number of communities $N_c$ in $G$ and the number of communities $N_c'$ in $G'$, thus:

**IL**$(G, G', m_6) = |$ **diff**$(m_6, m_6')|$     (2)

The value obtained from equation (2) can be normalized in order to compare between different benchmark datasets.

## 4 METHOD DESCRIPTION

We present a method which is based on selecting the $k$ most similar nodes and then perturbing them to make them identical, either by clustering or by modification. The similarity metric approximates an isomorphism matcher and also takes into account the degrees of the neigh-bors of the reference node. The local neighborhood sub-graph matching method herein described, has been presented as a European Patent application[22].

The algorithm operates in two phases: (i) a 'training' phase in which the weights are learned for the distance metric from samples and (ii) a 'runtime' phase which processes the complete dataset, matching nodes using the trained distance metric, and anonymizing them to obtain k-anonymity. Two main anonymization methods are used: clustering and modification. To these methods we allow two variants: with restrictions and without restrictions. The methods without restrictions are designated as having a 'global' search capacity. Finally, to the restricted methods, we implement two search variants: 'local1' and 'local2'. The methods and variants are summarized in Table 1, and will be explained in detail in the following Sections.

TABLE 1
SUMMARY OF ANONYMIZATION METHODS

|  | Name | Restrict | Search type | | |
| --- | --- | --- | --- | --- | --- |
|  |  |  | Global | Local1 | Local2 |
| **Clustering** | clust_r_l1 | Yes | No | Yes | No |
| **Clustering** | clust_r_l2 | Yes | No | No | Yes |
| **Clustering** | clust_g | No | Yes | No | No |
| **Modific.** | modif_g | No | Yes | No | No |
| **Modific.** | modif_r_l2 | Yes | No | No | Yes |

## 4.1 Definition of local neighborhood subgraph

In order to clearly define what we understand as a 'local neighborhood sub-graph, we will refer to Fig. 2. In Fig. 2 we see sub-graph $G_1$ whose reference node $N_{10}$ has three immediate (one hop neighbors) designated as $N_{11}$, $N_{12}$ and $N_{13}$. Neighbor $N_{11}$ has a degree of four, comprised of two neighbors that are internal to the sub-graph and two neighbors which are external to the sub-graph. We note that the external neighbors do not form part of the local neighborhood sub-graph and are only considered in order to calculate the degree of the neighbor nodes of $N_{10}$. The internal and external degrees of neighbors $N_{12}$ and $N_{13}$ are defined in a similar manner. Hence, a local neighborhood sub-graph around a given target node is defined as being comprised of the immediate neighboring nodes of said target node and their interconnections (links) between each other within the sub-graph and with the target node.

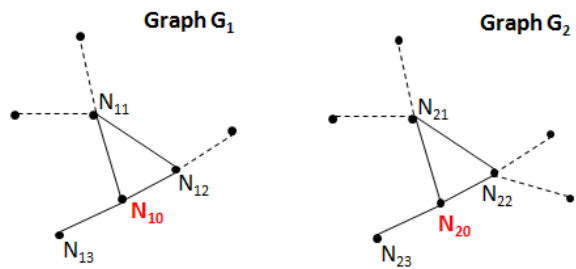

Fig. 2. Two local neighborhood subgraphs

## 4.2 Distance metric for similarity based selection

In the present work, we define a hybrid approach which adopts some aspects of the partitioning approach [1], and other aspects of the cost function approach [2][11][12]. We effectively obtain generalized partitions which minimizes information loss by using a similarity function to choose node pairs (and their local neighborhoods) for anonymization whose distance is a minimum. Hence, the information loss caused by their perturbation will be proportionately minimized.

We recall that the objective of identifying similar subgraphs (local neighborhoods) is in order to cluster or modify nodes as a mechanism to provide *k* anonymity. Each subgraph is considered as being the one hop neighborhood around a given node which is to be aggregated. Hence, we wish to aggregate node pairs which are as similar as possible, in terms of a given set of descriptive characteristics.

In order to calculate the similarity between two node neighborhoods, computation cost is a key consideration. Hence we have chosen a similarity metric which calculates a distance based on subgraph characteristics which can be pre-calculated. The subgraph characteristics are: degree of the reference node $D_R$; number of edges in the subgraph $N_E$, clustering coefficient CC, normalized average degree of adjacent nodes $AD_{AN}$, normalized standard deviation of degree of adjacent nodes $SD_{AN}$. The first three characteristics are designed to reflect the internal structure of the subgraph, whereas the last two characteristics reflect a key characteristic of the neighbors (their degree), which effectively considers the neighborhood one hop further out (from the reference node as starting point). We observe that in order to perform the calculation, all values are normalized against the maximum and minimum corresponding values in the complete graph.

The objective of the training step is to optimize the weights and obtain a function which models an isomorphism matcher/ neighbor degree matcher, but which has a much lower runtime computation cost, once we have trained the weights and pre-calculated the necessary statistical values.

In order to capture isomorphic and neighborhood characteristics, in a preprocessing step we trained the weights of the distance metric (there is one weight for each characteristic) using simulated annealing and a special isomorphism matcher to calculate the fitness value. We have used the VF2 isomorphism matcher [23], given that it is one of the fastest algorithms currently available, and is now widely used in the graph mining community. We have adapted VF2 so that it also returns a score of how well the respective neighborhood nodes maintain their degree values. We call this matcher VF2-D.

**Precision:** in terms of the subgraph characteristics, after running 10 fold training, we obtained 100% test accuracy on the reference node degree value and clustering coefficient of the subgraph at one hop, 94% correlation with the neighbor degrees and 97% correlation with the number of internal edges of each neighbor. On average, for the test datasets, approx. 70% of the top *k-1* nodes ranked by the distance function were isomorphisms.

### 4.3 Graph alteration operators – clustering and modification

In order to compare the relative performance of the restricted and non restricted approaches, we have used two of the most common state of the art techniques in the literature: (i) node modification and (ii) node clustering. We note that all the methods use the same node matching function, which is described in Section 4.2.

**Node modification:** for this method we have implemented a technique based on node addition and edge addition/ deletion which obtains k-anonymity using a cost function based on the expected perturbation. This method is similar to the one presented by Zhou in [2]. Due to the unavailability of the original code, we have programmed and tested our own version. The implementation uses the distance measure (see Section 4.2) as the cost function, and selects nodes for matching in descending order of degree, as indicated in [2]. Also, when we add a node to increase the degree, we choose them smallest degree first, again following the guidelines of [2]. Finally, edges are added to obtain the same internal degree sequences and minimize the difference between the respective sub-graph clustering coefficients. Hence, for node modification, two subgraphs $SG_1$ and $SG_2$ are considered equal when, for the reference node $g_1$ of $SG_1$ and the reference node $g_2$ of $SG_2$: degree($g_1$) = degree($g_2$) , num_edges($SG_1$) = num_edges($SG_2$) and internal_degree_sequence($SG_1$) = internal_degree_sequence($SG_2$). As mentioned in the introduction and in the privacy model definition (Section 5.2) we apply adversary queries based on structural similarity of node neighborhoods [1] rather than on isomorphic properties[4]. Hence, this equality criterion is adequate for both the type of adversary queries we consider in the current paper, and in order to compare the relative performance of the different methods under the same conditions.

Two versions are implemented: the first has no restrictions so it can choose nodes to match anywhere in the graph. We call this 'modif_g'; the second is restricted by the community, hub and bridge nodes, and is called 'modif_r_l1'. The restricted version will be described in Section 4.4.

**Node clustering:** for this method we have implemented a node aggregation method which groups the nodes into supernodes each of which contains at least k and at most *2k-1* of the original nodes. An optimum clustering is obtained by using a similarity function (see Section 4.1) to pair the most similar nodes for aggregation for each *k* value. Hence, for each node in the graph, the *k-1* most similar nodes will be identified as these nodes will unified into one supernode. If there are *2k* or more identical nodes (that is, taking into account that some nodes will already be identical in the graph), they will be grouped in supernodes each containing at least *k* nodes and at most *2k-1*. Three versions are implemented: the first has no restrictions so it can choose nodes to match anywhere in the graph. We call this 'clust_g'; the second is restricted by the community, hub and bridge nodes, as described in Section 4.4 and tries to match nodes in as local a neighborhood as possible. We call this 'clust_r_l1'; finally, the third is restricted in the same manner as the second, but can match nodes anywhere within a given community. We call this 'clust_r_l2'. The restricted versions will be described in Section 4.4.

Hence we have two modification methods, 'modif_g' and 'modif_r_l1' and three clustering methods, 'clust_g', 'clust_r_l1' and 'clust_r_l2'.

## 4.4 Search restriction variants: 'global', 'local1' and 'local2'

We consider three search strategies: (i) 'global' in which nodes can be searched for and matched anywhere in the graph; (ii) 'local1' in which node search and matching is restricted to the same community and excludes top hub and bridge nodes. This method further restricts by trying to find the best match which is also locally the closest to the reference node; (iii) 'local2' in which node search and matching is restricted to the same community and excludes top hub and bridge nodes. In contrast to 'local1', 'local2' can search for nodes anywhere within the same community.

**Local1:** searches in an expansive manner around the immediate neighborhood of the reference node, within the community to which the reference node belongs. It looks for node matches whose similarity distance is within a given threshold θ (assigned as the average similarity for all node pairs in the complete graph). **Local2:** searches for the best match of a given reference node, anywhere in the community to which the reference node belongs.

For the restricted methods we initially execute a "Community Structure" algorithm to partition the complete graph into "communities". We use Blondel's algorithm, known as the Louvain Method[21]. We also identify the top 12% percentile hub nodes and 10% percentile bridge nodes by calculating their corresponding metrics. The percentile values were chosen by empirical study of the metric distributions. In practice, these top percentile proportions tend to represent a small number of key nodes in the graph.

## 4.5 Pseudo-code of method

In this section we define the main procedures which comprise the approach: "Precalculate", "Train" and "Run" (the latter calls each of the five methods).

**Procedure Main**

*Input:* original graph G = (V, E), anonymization level *k* *Output:* anonymized graph G')

1. *Precalculate*
2. Calculate statistics for each local area subgraph $SG_1 \ldots SG_n$
3. Calculate bridge and hub metrics
4. Calculate communities $c_1 \ldots c_i$ using Louvain method
5. *Train*
6. For each sample *s*, apply simulated annealing process to find
   optimum weights for distance function
7. *Run*
8. Let B be the set of bridge nodes b above the bridge
   percentile threshold
9. Let H be the set of hub nodes h above the hub
   percentile threshold
10. Let *k* be the privacy level

11. FOR EACH (g) ϵ (G) , g ∉ B , g ∉ H
12.    Let $c_i$ be the community to which node g belongs
13.    Define $SG_{g1}$ as the local neighborhood sub-graph for g
14.    Call methods
15.     **Clustering methods:**
16.      Find *k-1* nodes most similar to *g* ‡
17.       **clust_r_l1**(graph $SG_{g1}$, $c_i$, *k*) // local 1, restricted
18.       **clust_r_l2**(graph $SG_{g1}$, $c_i$, *k*) // local2, restricted
19.       **clust_r**(graph $SG_{g1}$, *k*)    // global, unrestricted
20.     Aggregate the *k* local neighborhood subgraphs $SG_g$ and [$SG_{g2}$ ... $SG_{gk}$] by calling
      **Aggregate**(vector of subgraphs[$SG_{g1}$ ... $SG_{gk}$])
21.    **Modification methods:**
22.     Find *k-1* nodes most similar to *g* ‡
23.      **modif_r_l2**(graph $SG_{g1}$, $c_i$, *k*) // local2, restricted
24.      **modif_r**(graph $SG_{g1}$, *k*)    // global, unrestricted
25.     Modify the *k* local neighborhood subgraphs [$SG_{g2}$ ... $SG_{gk}$] to make them the same as $SG_g$ by calling
      **Modify**(vector of subgraphs[$SG_{g1}$ ... $SG_{gk}$])
26. END FOREACH
   ‡Each method returns the best *k-1* matches [SGg2 ... SGgk] that comply with restrictions

## 5 METRICS FOR INFORMATION LOSS, PRIVACY LEVEL AND RISK OF DISCLOSURE

In this Section we give our definitions for information loss and risk of disclosure. Information loss is defined in the habitual manner, as the change in correlation between each variable's data in the original file and the perturbed file. For risk of disclosure we define a set of candidate anonymity queries, similar to those of Hay[1].

### 5.1 Information Loss

We use six metrics in order to evaluate information loss. The first three are basic graph statistics (degree, clustering coefficient and average path length), and the last three are related to the community structure of the graph (hub metric (HITS), betweenness centrality and number of communities). The distribution of each variable in the original data file is compared (correlated) with that of the same variable in the perturbed file, and the deviation from 1 is the information loss.

    **inf loss$_1$** degree
    **inf loss$_2$** clustering coefficient
    **inf loss$_3$** average path length
    **inf loss$_4$** hub value (HITS)
    **inf loss$_5$** bridge value (betweenness centrality)
    **inf loss$_6$** number of communities*

\* As calculated by Louvain method

We have implemented the 'cc', 'apl' and 'bridging centrality' algorithms in Java. In the case of the 'apl' (average path length) statistic, we have used Dijkstra's algorithm[24]. In the case of hub value (HITS) and betweenness centrality, we have calculated these using the Gephi software[25].

   **Bridge nodes.** These are nodes which may not necessary have a high degree but which are "strategically" placed between nodes such that they form key part of the graph structure. That is, their removal would cause a major disruption to the graph structure. It can be considered in terms of the number of critical paths which go through it, from/to other nodes (this is known as "betweenness centrality". We use the measure published by Hwang et al. in [20], called "bridging centrality", and which is very effective in distinguishing bridge nodes from hub nodes.

**Hub metric (HITS hub).** A hub node is characterized by having a large number of direct connections to other nodes. In order to quantify the hub value of a node, we have used the popular HITS algorithm, as defined by Kleinberg in [19].

**Communities.** The community partitioning is a key characteristic of the graph that we wish to maintain. We measure information loss by the number of communities into which the graph is partitioned, as calculated by the Louvain method [21].

### 5.2 Definition of Privacy for clustering and modification methods

The objective of anonymization is to obtain a given anonymity level of $k$, as stated in Section 3. As described in Section 4, the clustering algorithm is given the parameter $k$ and produces a graph consisting of supernodes which contain a minimum of $k$ and a maximum of $2k-1$ basic nodes. If a supernode reaches a size of $2k$ nodes, it will be divided into two supernodes, each containing $k$ nodes. Nodes are grouped based on similarity using the distance metric described in Section 4.2. Hence, nodes are grouped into partitions (each containing between $k$ and $2k-1$ nodes), so that an adversary will be unable to distinguish between the nodes in a partition. The probability that an adversary successfully re-identifies a node will be between $1/k$ and $2k-1$, multiplied by the number of supernodes created for the given reference node (we note that supernodes are also created using identical nodes). For the modification algorithm, we have implemented an approximation of Zhou's method[2] which, for a given node, modifies $k-1$ other nodes to make them the same (using our distance based similarity metric). That is, for each node there will be $k-1$ other nodes with the same degree, number of edges in the one hop subgraph, and same clustering coefficient (that is, the connectivity between neighbors). Hence, the probability of an adversary re-identifying a node will be at least $1/k$. We note that nodes which are already identical will not be modified and there will probably be nodes in the graph which already have more than $k$ identical nodes (especially the low degree nodes).

As commented in Section 3, we employ an anonymity model such that a graph satisfies *k-candidate* anonymity if for every structural query over the graph, there exist at least $k$ nodes that match the query (the structural queries are defined in the following section).

### 5.3 Adversary Knowledge - structural queries

We follow similar lines to Hay[1] in terms of what the attacker knows or can deduce from the graph. Firstly, we consider *vertex refinement*, as defined by Hay in [1]. Then we consider *subgraph queries*, *hub fingerprint* and a new attack we define for the first time in this paper, which we will call *bridge fingerprint*.

**Vertex refinement[1]:** $H_1(x)$ returns the degree of $x$, $H_2(x)$ returns the multi-set of each neighbors' degree, and so on. In general, $H_i(x)$ returns the multi-set of values which are the result of evaluating $H_{i-1}$ on the set of nodes adjacent to $x$. We consider up to two levels of query, $H_1$ and $H_2$ as defined in [1].

**Subgraph queries[1]:** number of edge facts an adversary needs to know in order to identify a subgraph around a target node. The subgraph query $SG(x)$ returns the number of edges in the subgraph of a node $x$ with its immediate neighbors (1 hop).

**Hub fingerprint[1]:** a hub fingerprint query $F_i(x, HB)$ gives a list of the shortest paths from node $x$ to each of the hub nodes defined in the vector HB. Hay defines HB as the five highest degree nodes for the Enron datasets and the ten highest degree nodes for the Hep-th and Net-trace datasets. Following Hay[1] we assume that the value $i$ designates the maximum distance of visible hub connections. If the shortest path to a hub exceeds the 'visibility horizon' then the distance is assigned value zero (open world assumption). Hence, query $F_1(x, HB)$ returns the list for $x$ with a visibility horizon of 1 hop, and $F_2(x, HB)$ returns the list for x with a visibility horizon of 2 hops. As an example, consider $F_2(x, HB)$, HB = {$a, b, c$} and a resulting distance vector of {2, 2, 0}. This means that node $x$ is at distance 2 from hubs '$a$' and '$b$', and at a distance greater than 2 (beyond the visibility horizon) from hub '$c$'.

**Bridge fingerprint:** a bridge is a node which may not have a high degree, but acts as a con-nector between a high number of nodes. We can define two types of bridge: (i) local bridges which connect hubs, that is, the shortest path between two or more hubs is through the bridge node. We can also define a local limit, such as 1 or 2 hops, to the subgraph considered; (ii) global bridges through which many shortest paths pass from all over the network. Given that global bridges are much more difficult to identify, we will choose local bridges as the fingerprint, given that an attacker will find it easier to identify the hubs and the nodes which act as bridges between in a local neighborhood. Thus we define a bridge fingerprint query $F_i(x, BR)$ which gives a list of the shortest paths from $x$ to each of the bridge nodes defined in the vector BR, with a visibility horizon of $i$.

## 6 EXPERIMENTAL SETUP

In this Section we describe the datasets used and the *modus operandi* of the experiments.

### 6.1 Datasets

We have used the Ca-HepTh[26], Enron [27] and WikiVote [28] datasets for empirical testing. These datasets offer distinct statistical characteristics and are widely used in the graph privacy literature, which allows other researchers to compare results. In the remainder of the paper, we will refer to these datasets as 'hepth', 'enron' and 'wikivote', respectively. The 'hepth' and 'wikivote' datasets were taken directly from the Stanford Large Network Dataset Collection (SNAP) website (available at http://snap.stanford.edu/data/ ). In the case of the 'enron' dataset, we processed the data ourselves from the mysql dump file available at http://www.isi.edu/~adibi/Enron/Enron.htm. In Table 2 we see the basic statistics for the three graphs datasets. We note the relatively low average clustering coefficient for the 'wikivote' dataset, and the relatively low average degree, high average path length, high diameter and large number of communities for the 'hepth' dataset.

TABLE 2
SUMMARY OF GRAPH STATISTICS FOR THE TEST DATASETS*

|  | hepth | enron | wikivote |
|---|---|---|---|
| **#Nodes** | 9877 | 10630 | 7115 |
| **#Edges** | 51971 | 329674 | 103689 |
| **Avg. degree** | 5.259 | 31.014 | 28.324 |
| **Clust. coef.** | 0.471 | 0.384 | 0.141 |
| **Avg. path length** | 5.945 | 3.160 | 3.247 |
| **Diameter** | 18 | 9 | 7 |
| **Nº communities** | 472 | 51 | 40 |

*All statistics have been calculated using the Gephi software on the original datasets used.

### 6.2 Benchmark methods and Experiments

Five contrasting methods are benchmarked, which have been previously described in Section 4. These consist of two modification methods, 'modif_g' and 'modif_r_l2' and three clustering methods, 'clust_g', 'clust_r_l1' and 'clust_r_l2'. We apply the given methods to the datasets stated in Section 6.1, to generate four anonymized graph datasets, corresponding to k=2, 4, 8 and 16. Then to each of these datasets and to the original dataset we apply calculate the graph statistics defined in Section 5.1 in order to compute the information loss. Finally, to the same datasets we apply the five adversary queries defined in Section 5.3 in order to calculate the risk. This enables us to evaluate the relative performance of the five perturbation methods, from different perspectives: information loss metrics, adversary queries, dataset characteristics, clustering versus modification, and restrictions versus no restrictions.

## 7 EMPIRICAL TESTING AND RESULTS

In this Section we present the results for information loss and adversary success for the different methods, metrics and datasets.

### 7.1 Information Loss Vs anonymization (k) level

For the metrics of Fig. 3 (degree, cc and apl) and the first two metrics of Fig. 4 (hub and bridge), the information loss is quantified by first calculating the graph metrics for the different graph datasets corresponding to k=0, k=2, k=4, k=8 and k=16. Then we correlate the value for each metric for the k=0 dataset with each of the other datasets (k=2, …). The difference between the correlations is then interpreted as the information loss. For the third metric of Fig. 4, 'number of communities', the absolute values are plotted and compared. In Figs. 3 and 4 we depict the information loss for progressively increasing anonymization levels: that is, increasing values of *k*. In Fig. 3 we correlate the distribution of the characteristic (for example, degree) in the original dataset with the distribution of the characteristic in the anonymized dataset (for example, *k*=2). In the first row of Fig. 3 we observe that, for degree, all methods follow a similar trajectory. 'modif_g' displays the smallest change for increasing values of *k*, and hence it is the method with the least information loss for the 'degree' metric. The trends are very similar for each of the three datasets. In the second row of Fig. 3 (clustering coefficient metric) we again see that 'modif_g' suffers the least information lost whereas 'clust' shows the most. We see the other three (restricted) methods with an intermediate performance, the relative position varying slightly depending on the dataset. In the third row of Fig. 3 (average path length metric) we see a somewhat different scenario to degree and clustering coefficient. For this metric, the restricted methods display the least information loss. We also observe that the 'Wikivote' dataset suffers a higher information loss in general and the atypical behavior of the 'clust' method for the 'HepTh' dataset.

Now turning to Fig. 4, the first row shows the information loss for the 'hub' metric. For the 'enron' dataset, 'clust_r_l1' and 'clust_r_l2' gave the best results, together with 'clust_g'. For the 'Wikivote' and 'hepth' datasets, 'modif_r_l2' gave the least information loss, followed by 'clust_r_l1' and 'clust_r_l2'. The 'Wikivote' dataset suffers the highest information loss. In the second row of Fig. 4 we see the information loss for the 'bridge' statistic. We observe that the 'wikivote' dataset has again given the greatest information loss, with the value losing practically all its correlation for all methods with the exception of 'modif_r_l2'. This is similar to the result for the average path length (apl) and the same dataset. This is logical if we consider that the bridges 'hold together' the different regions of the graphs, hence a loss to the bridges will greatly affect the traversal distances across the complete graph. We recall that one of the measures we have taken to 'protect' the graph topology is to exclude from aggregation the top 10% percentile, for the restricted methods. In the case of the 'hepth' dataset this seems to have worked relatively well, with an information loss of between 40 and 50% for *k*=16. In the case of the 'enron' dataset, the information loss also compares favorably for the restricted methods, with respect to the non-restricted methods which have no protection for bridges. The reasons for the poorer performance for 'wikivote' must be topological, and could be related to a higher proportion of significant bridging nodes which are however not in the top 10% percentile.

The final row of Fig. 4 shows the raw values of a graph metric which is specifically related to the community structure of the graph: number of communities. We are particularly interested in this feature and have expressly implemented restrictions on the perturbation to protect the overall 'community structure' of the graph, as described previously. In the third row of Fig. 4 we see the effect of increasing *k* on the number of communities. We see that the restricted methods give much better results than the non-restricted ones, in terms of the change in the number of communities, for increasing levels of *k*. We see in all cases that the 'clust' (unrestricted) method displays the highest information loss. For the 'Wikivote' dataset, the 'modif_g' method displayed an atypical behavior for k=2. We take note that the 'hepth' dataset has approx. 10 times as many community partitions as the other two datasets. In summary of these results, we can conclude that the measures taken to protect the community structure have worked as expected, minimizing the change in the corresponding key characteristic.

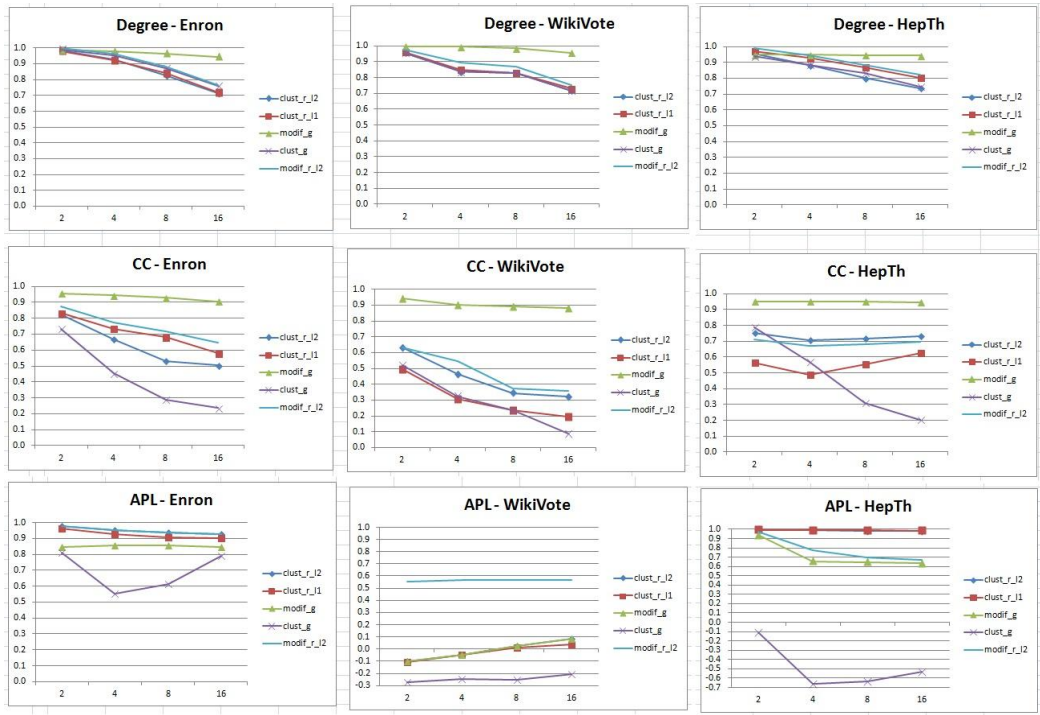

Fig. 3. **Information Loss: graph characteristics**. Effect of anonymization on three key graph characteristics for three different datasets. The figures show the degree of correlation (y-axis) between the original data and perturbed data for different values of *k* (x-axis). Five perturbation methods are shown: 'clust_r_l2', 'clust_r_l1', 'clust_g', 'modif_g' and 'modif_r_l2'.

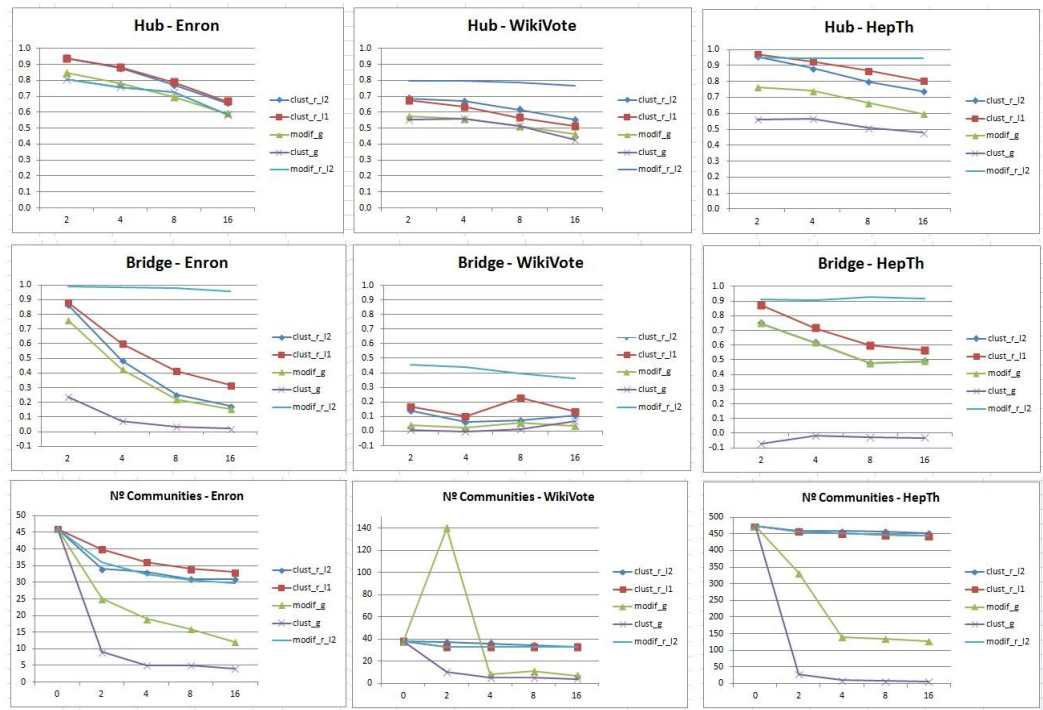

Fig. 4. **Information Loss: community characteristics**. Effect of anonymization on two three restricted characteristics (hub and bridge metrics, and number of communities) for three different datasets. The figures show the raw data values (y-axis) for different values of *k* (x-axis). Five perturbation methods are shown: 'clust_r_l2', 'clust_r_l1', 'clust_g', 'modif_g' and 'modif_r_l2'.

In Tables 3 and 4 we see a quantified summary of the relative performance of the methods, in terms of the number of times they were first, second, third or fourth best in each of the 18 cases shown in Figs. 3 and 4. If two methods gave a tie, for example, for first position for a given case, both methods were awarded one point for first position.

We conclude that 'modif_r_l2' is the clear overall winner, followed by 'clust_r_l1'. However, for the 'degree' and 'cc' metrics, 'modif_g' (an unrestricted method) gave the least information loss. Hence, we can conclude that the restrictions (community, hub and bridge) did not optimally mitigate information loss for the 'degree' and 'clustering coefficient' metrics. However, as can be seen from Table 3, they did so for the other four metrics. Thus, we see that some of the relative performances are dataset and metric dependent.

TABLE 3
INF. LOSS: RELATIVE PERFORMANCE OF METHODS BY METRIC.

|        | clust_r_l2 | clust_r_l1 | modif_g | clust_g | modif_r_l2 |
|--------|------------|------------|---------|---------|------------|
| degree | 3.7        | 3.0        | 1.0     | 3.3     | 2.0        |
| cc     | 3.0        | 3.7        | 1.0     | 5.0     | 2.3        |
| apl    | 2.0        | 2.0        | 2.7     | 4.0     | 1.3        |
| hub    | 2.0        | 2.0        | 3.7     | 4.0     | 1.3        |
| bridge | 3.0        | 2.0        | 3.7     | 4.7     | 1.0        |
| NC     | 1.3        | 1.0        | 2.3     | 3.3     | 1.3        |
| Avg.   | 2.5        | 2.3        | 2.4     | 4.1     | 1.6        |

TABLE 4
INF. LOSS: RELATIVE PERFORMANCE OF METHODS BY DATASET.

|          | clust_r_l2 | clust_r_l1 | modif_g | clust_g | modif_r_l2 |
|----------|------------|------------|---------|---------|------------|
| enron    | 2.5        | 2.0        | 2.5     | 3.8     | 1.7        |
| wikivote | 2.5        | 2.7        | 2.3     | 4.2     | 1.3        |
| hepth    | 2.5        | 2.2        | 2.3     | 4.2     | 1.7        |
| Avg.     | 2.5        | 2.3        | 2.4     | 4.1     | 1.6        |

## 7.2 Risk (adversary information) Vs anonymization (k) level

The risk is quantified by applying five different adversary queries, as we have previously described in Section 5.3. The risk is measured in terms of candidate set sizes, following the guidelines of Hay[1]. That is, the highest risk exists for nodes for the lowest candidate set size (=1), whereas the lowest risk exists for nodes for the highest candidate set size. Throughout the following text we will use the term 'bucket' as a synonym for 'candidate set'.

In Figs. 5 and 6 the risk is plotted for each of the adversary queries, for each dataset and for increasing values of *k*. For space restrictions, only the lowest risk candidate set (bucket) is shown for each adversary query. However, in the text we comment the proportions in the other buckets, whenever significant. The proportion of nodes in the lowest risk bucket is a key indicator of risk and was the bucket which best characterized the adversary queries and methods.

The buckets for adversary queries 1, 2, 4 and 5 were defined with the following frequencies: '=1', '2-4', '5-10', '11-20' and '>20'. For adversary query 3 (edges), the following buckets were defined: '=1', '2-10', '11-100', '101-1000' and '>1000'.

We also make clear that the candidate sets are based on the number of nodes returned by the adversary queries for the anonymized graphs. This gives a vision from the viewpoint of the adversary. We observe that, for the clustering methods, each node (supernode) returned will contain a minimum of *k* and a maximum of *2k-1* basic nodes (see Section 5.2). Hence, for k=16, every aggregated (super) node contains between 16 and 31 elemental nodes of the original graph. Therefore, although to the adversary it appears that s/he has found a unique node in the graph, this node really contains between 16 and 31 elemental nodes. Internally (not published), we know how many basic nodes each supernode contains and this is used to place it in the corresponding frequency bucket. In the case of the modification methods, for each node there will be at least *k-1* identical nodes in the graph. Hence, when an adversary query returns one node, its frequency will be equal to the number of nodes, which will be at least *k-1*, labeled as being identical to it in the graph. In likelihood terms, if the adversary is trying to identify a specific node (for example, corresponding to user 'john'), and the query returns a single node, then the chances of that node being that of 'john' will be at least *1/k*.

### 7.2.1 Adversary query 1: vertex refinement $H_1(x)$

Fig. 5 (row 1) shows the trends for the different candidate sets, datasets and values of *k*, for the first adversary query, vertex refinement $H_1(x)$. We recall that this query simply returns the degree of a given node. If we first consider Fig. 5 (row 1) in terms of the proportion of nodes in each candidate set, we observe for all original datasets (*k=0*) that the great majority (90%) of the degree values are in the highly frequent candidate set ('>20', low risk), and the remaining 10% are distributed through the other, higher risk candidate sets, '=1', '2-4', '5-10' and '11-20'.

In the case of the bucket '11-20', all methods followed a similar trend with the exception of 'clust_g' which followed a markedly upward gradient, giving a significantly higher relative proportion of nodes in the bucket '11-20', with respect to the other methods. Also, 'modif_g' displayed a slightly different tendency, with a relatively lower proportion of nodes in the bucket for the 'enron' and 'wikivote' datasets. We note that for k=16, there will be zero nodes in the '=1', '2-4' and '5-10' buckets.

If we now look at Fig. 5 (row 1), we see that all methods follow a similar trend, with the exception of 'clust_g' which shows a lower proportion of nodes in the '>20' bucket, with respect to the other methods.

### 7.2.2 Adversary query 2: vertex refinement $H_2(x)$

Fig. 5 (row 2) shows the trends for the different candidate sets, datasets and values of *k*, for the second adversary query, vertex refinement $H_2(x)$. We recall that this query returns the degrees (in a vector) of each of the immediate neighbors of a given node. For the bucket '11-20', the proportion remained low until k=16, which is when all the nodes from the previous bucket ('5-10') appeared in bucket '11-20'. In terms of the methods, all methods follow a similar trend for all three datasets, with the exception of bucket '>20' and dataset 'hepth'. In this last case the 'clust_g' method display a progressively decreasing proportion in contrast to all other methods. Also, we can see that the 'modif_r_l2' and 'clust_r_l1' methods display a slightly higher relative proportion of nodes for the 'wikivote' and 'hepth' datasets and the '>20' bucket.

### 7.2.3 Adversary query 3: subgraph edge query $SG(x)$

Fig. 5 (row 3) shows the trends for the different candidate sets, datasets and values of *k*, for the third adversary query, subgraph SG(x). We recall that this query returns the number of edges in the subgraph formed by the immediate neighbors of a given node. Firstly we note that the candidate set size ranges are distinct from those of the degree (Fig. 5, row 1) and neighbor degrees (Fig. 5, row 2). This is because the magnitude of the value 'number of edges' for the subgraph considered is much greater than the degree values. Hence the set sizes reflect this, with the biggest set size (lowest risk) set to '>1000'. The set sizes were assigned by observing the distributions for each dataset of the number of edges. We were also initially guided by the ranges used by Hay in [1] for the edge fact.

In general we see similar trends for the different datasets and buckets, with the following exceptions: for the 'hepth' dataset the 'clust_g' method displays a higher relative proportion of nodes in bucket '101-1000' and a much lower relative proportion of nodes in bucket '>1000', with respect to the other methods; also, the 'modif_g' method shows a distinctive trend in several cases, displaying a lower proportion for the '101-1000' bucket.

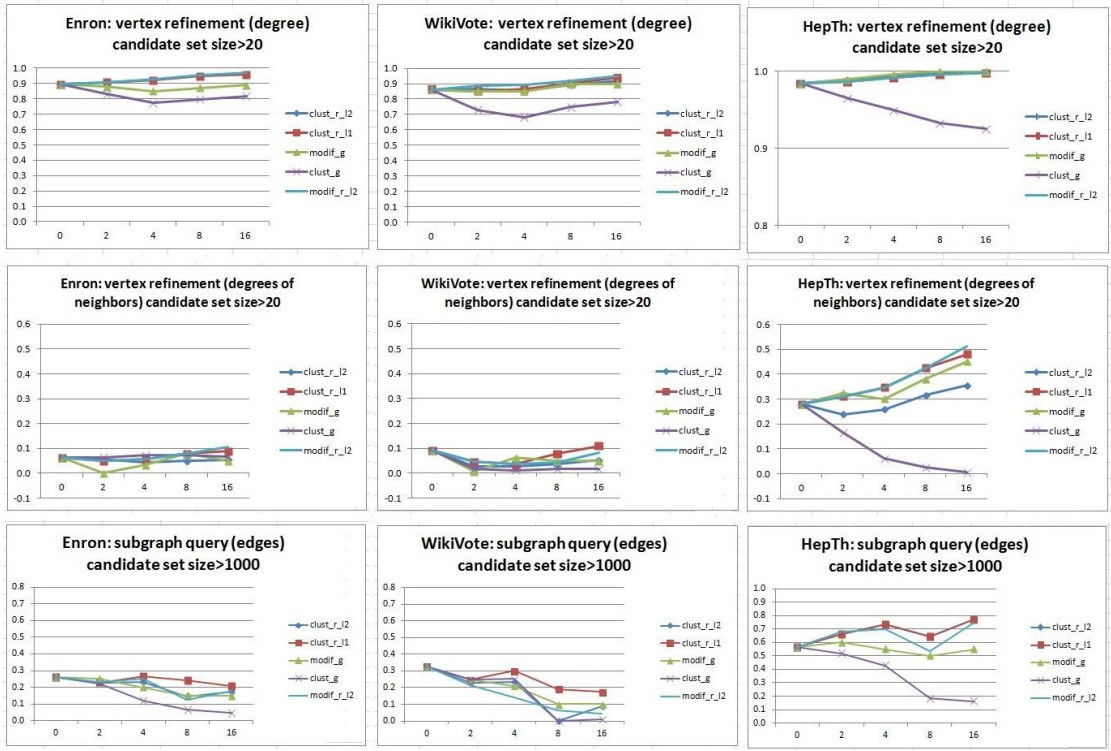

Fig. 5. **Risk, Adversary Queries on degree (vertex refinement $H_1(x)$), degrees of neighbors (vertex refinement $H_2(x)$ and edges (subgraph query $SG(x)$).** Effect of anonymization on risk for candidate sets '>20' (degrees and degrees of neighbors) and '>1000' (edges), for three different datasets. The figures show the percentage of nodes (y-axis) in a given candidate set bucket for different values of $k$ (x-axis). Three clustering based and two modification based perturbation methods are compared.

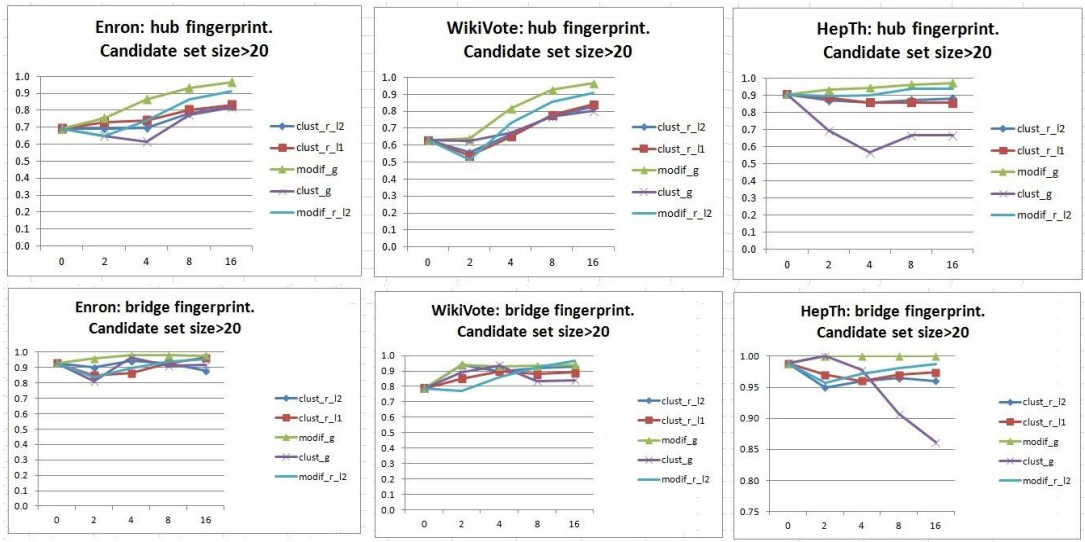

Fig. 6. **Risk, Hub and Bridge Fingerprints with visibility horizon of 2**. Effect of anonymization on risk for candidate set '>20', for three different datasets. The figures show the percentage of nodes (y-axis) in a given candidate set bucket for different values of $k$ (x-axis). Three clustering based and two modification based perturbation methods are compared.

### 7.2.4 Adversary query 4: hub fingerprint $F_2(x, H)$

Fig. 6 (row 1) shows the trends for the different candidate sets, datasets and values of $k$, for the fourth adversary query, hub fingerprint (See Section 5.3). We recall that this query returns a vector of the shortest path length to a fixed set of 10 top hubs in the graph. The 'hub' value for each node is quantified by calculating the HITS 'Hub Update Rule' metric, as commented in Section 5.1. With respect to the original datasets ($k=0$), we observe that the majority of the vector frequencies are in the '>20' candidate set, initially containing approx. 69% of the nodes for 'enron', 63% for 'wikivote' and 92% for 'hepth'. As a result of anonymization up to k=16, in general we see an increase in the '>20' low risk set. In terms of the methods, for bucket '11-20', 'modif_g' displayed the lowest relative proportion. With reference to Fig. 6 (row 1), for bucket '>20', 'modif_g' shows the highest relative proportion whereas 'clust_g' displays the lowest or equal lowest.

### 7.2.5 Adversary query 5: bridge fingerprint $F_2(x, B)$

Fig. 6 (row 2) shows the trends for candidate set '>20', datasets and values of $k$, for the fifth and final adversary query, bridge fingerprint. We recall that this query returns a vector of the shortest path length to a fixed set of 10 top bridges in the graph. The 'bridge' value for each node is quantified by calculating the 'bridge centrality' metric, as commented in Section 5.1.

With respect to the original datasets ($k=0$), we observe that the majority of the vector frequencies are in the '>20' candidate set. This bucket initially contains approx. 92% of the nodes for 'enron', 80% for 'wikivote' and 99% for 'hepth'. The 'hepth' has the most 'extreme' candidate set membership in which practically all the nodes are in the '>20' set. This distribution is somewhat distinct from the 'hub fingerprint' query, which had approx. 20% less nodes in the '>20' set and approx. 40% of the remainder distributed among the mid-range sets, for the 'enron' and wikivote' dataset. In terms of the methods, we see a similar behavior for the different methods, with the following exceptions: 'clust_g' has a higher proportion of nodes in bucket '11-20' for k=16 and the wikivote and hepth datasets; 'clust_g' also has a progressively lower proportion of nodes for the '>20' bucket; 'modif_g', on the other hand, tends to show a slightly lower percentage for the '11-20' bucket and a slightly higher one for the '>20' bucket, relative to the other methods.

### 7.2.6 Summary and synthesis of the analysis of the adversary query results (Sections 7.2.1 to 7.2.5)

In this Section we will present an overall picture of the results of the different adversary queries, taking into account the detailed analysis which we have already seen in Sections 7.2.1 to 7.2.5. In order to synthesize the results in a quantitative manner, we will rank the methods and datasets, in terms of their performance for increasing values of $k$ and adversary query type. We will use the candidate set with the highest number of candidates (lowest risk) as the benchmark. It is clear that the more candidates that fall into this category the better because they will have the lowest identification risk. We will follow the same evaluation method as we did for the information loss in Section 7.1.

TABLE 5
RISK: RELATIVE PERFORMANCE OF METHODS BY ADV. QUERY.

|  | clust_r_l2 | clust_r_l1 | modif_g | clust_g | modif_r_l2 |
|---|---|---|---|---|---|
| $H_1(x)$ | 1.0 | 1.0 | 1.7 | 2.7 | 1.3 |
| $H_2(x)$ | 3.3 | 1.3 | 4.0 | 4.7 | 1.7 |
| $SG(x)$ | 2.3 | 1.0 | 2.7 | 4.3 | 2.3 |
| $F_2(x, H)$ | 3.3 | 3.3 | 1.0 | 4.7 | 2.0 |
| $F_2(x, B)$ | 2.7 | 3.0 | 1.0 | 4.3 | 3.0 |
| **Rank** | 3 | 1 | 2 | 4 | 2 |

TABLE 6
RISK: RELATIVE PERFORMANCE OF METHODS BY DATASET.

|  | clust_r_l2 | clust_r_l1 | modif_g | clust_g | modif_r_l2 |
|---|---|---|---|---|---|
| **enron** | 2.4 | 2.0 | 2.4 | 4.2 | 2.0 |
| **wikivote** | 2.6 | 1.6 | 2.0 | 4.0 | 2.6 |
| **hepth** | 2.6 | 2.2 | 1.8 | 4.2 | 1.6 |
| **Rank** | 3 | 1 | 2 | 4 | 2 |

In Tables 5 and 6 we see a quantified summary of the relative performance of the methods, in terms of the number of times they were first, second, third or fourth best in each of the 15 cases shown in Figs. 5 and 6. If two methods gave a tie, for example, for first position for a given case, both methods were awarded one point for first position. In these terms we see that 'clust_r_l1" is the overall winner followed by 'modif_r_l2' and 'modif_g'. In summary, and with reference to Table 5, we observe that for the first three adversary queries (first three rows of Table 5), the restricted and clustering methods (excluding 'clust_g') gave the lowest risk. However, for the last two adversary queries (hub and bridge fingerprints), the non-restricted method 'modif_g' gave the lowest risk. This can be explained by taking into account that the restricted methods do not perturb the top percentile hubs and bridges.

## 8 SUMMARY AND CONCLUSIONS

We have presented and tested two main methods for perturbing a graph, based on node modification and node clustering, combined with a set of restrictions which mitigate the effect of the perturbation, and hence the information loss, on key graph characteristics: community structure, hubs and bridges. We have seen that in terms of risk, the restricted methods gave the best results for the degree, neighbor degree and edge adversary queries. On the other hand, the unrestricted method (modif_g) gave the best results for the hub and bridge fingerprint adversary queries. This has been commented in Section 7.2.6.

In terms of information loss, the best method varies depending on the metric used, and in some cases on the dataset characteristics. The 'modif_g' (unrestricted) method was best for the degree and clustering coefficient metrics, for all datasets; 'modif_r_l2' was best for the Hub and Bridge metrics and was also best for the APL metric with the exception of the WikiVote dataset; 'clust_r_l1' was best or tied best for the NC (number of communities) metric for all datasets and came second best for all metrics except 'degree' and 'cc'. Hence, we can conclude that the restricted methods gave lower information loss in general (4 out of 6 metrics) and lower risk for 3 out of 5 of the adversary queries. The modification based method 'modif_r_l2' gave the lowest overall information loss whereas 'clust_r_l2' gave the lowest overall risk, with 'modif_r_l2' a close second. A general summary can be seen in Table 7.

TABLE 7
GRAPH CHARACTERISTICS VS BEST PERFORMING METHODS

|  | Enron | HepTh | WikiVote |
|---|---|---|---|
| **Dataset characteristics** | High D, high CC, low NC, v. high ACS | High D, low CC, low NC, high ACS | Low D, high CC, high NC, low ACS |
| **Risk minimization** | 'clust_r_l1', 'modif_r_l2' | 'clust_r_l1', 'modif_r_l2' | 'clust_r_l1', 'modif_r_l2' |
| **Information loss minimization** | 'modif_g' (D, CC); 'modif_r_l2' (APL, Hub, Bridge); 'clust_r_l1' (NC) | 'modif_g' (D, CC); 'modif_r_l2' (APL, Hub, Bridge); 'clust_r_l1', 'clust_r_l2', 'modif_r_l2' (NC) | 'modif_g' (D, CC); 'modif_r_l2' (Hub, Bridge); 'clust_r_l2' and 'clust_r_l1' (APL); 'clust_r_l1', 'clust_r_l2', 'modif_r_l2' (NC) |

*D=degree, CC=Clustering Coefficient, NC=number of communities, ACS=average community size, APL=Average Path Length.

**ACKNOWLEDGMENTS**

This research is partially supported by the Spanish MEC (projects ARES CONSOLIDER INGENIO 2010 CSD2007-00004 -- eAEGIS TSI2007-65406-C03-02 -- and HIPERGRAPH TIN2009-14560-C03-01)..